\documentclass[aps,prl,showpacs]{revtex4}
\usepackage{amsmath}
\usepackage{graphicx}
\usepackage{bm}

\begin{document}

\title{Bound states of the s-wave Klein-Gordon equation with equal scalar
and vector Standard Eckart Potential}
\date{\today}
\author{Eser Ol\u{g}ar}
\affiliation{Department of Physics, Faculty of Engineering University of Gaziantep, 27310
Gaziantep, Turkey}
\email{olgar@gantep.edu.tr}
\author{Ramazan Ko\c{c}}
\email{koc@gantep.edu.tr}
\affiliation{Department of Physics, Faculty of Engineering University of Gaziantep, 27310
Gaziantep, Turkey}
\author{Hayriye T\"{u}t\"{u}nc\"{u}ler}
\email{tutunculer@gantep.edu.tr}
\affiliation{Department of Physics, Faculty of Engineering University of Gaziantep, 27310
Gaziantep, Turkey}

\begin{abstract}
A supersymmetric technique for the bound-state solutions of the
s-wave Klein--Gordon equation with equal scalar and vector
standard Eckart type potential is proposed. Its exact solutions
are obtained. Possible generalization of our approach is outlined.
\end{abstract}


\maketitle

\section{Introduction}

Relativistic quantum mechanics is required to obtain more accurate
results for the particle under a strong potential field. When we
consider this condition, a particle in the strong potential field
should be described by the Klein--Gordon equation and the Dirac
equation.$^{[1,2]}$ In order to analyse relativistic effects on
the spectrum of such a physical system, one may construct the
Klein--Gordon equation including adequate potentials and obtains
their solutions. In recent years, there have been many discussions
about the Klein--Gordon equation with various types of potentials
by using different methods to obtain the spectrum of the system.
Some authors have considered the equality of scalar potential and
vector potential in solving the Klein--Gordon equation as well as
the Dirac equations for some potential fields.$^{[3-7]}$ The
s-wave bound-state solutions are obtained in Refs.\,[5-8].
Similarly, the s-wave Klein--Gordon equation with vector potential
and scalar Rosen-Morse type potentials$^{[9,10]}$ has been treated
by the standard method,$^{[11]}$ the same problem with both the
vector and scalar Hulthen-type potentials have been discussed
analytically.$^{[12]}$ Energy spectrum of the s-wave
Schr\"{o}dinger equation with the generalized Hulthen potential
has been obtained by using the supersymmetric quantum mechanics
(SUSYQM) and supersymmetric Wentzel--Kramers--Brillouin (WKB)
approach.$^{[13]}$ The bound-state spectra for some physical
problems has been studied by the quantization condition and the
SUSYQM.$^{[14-16]}$

In this Letter, we construct a Klein--Gordon equation including
the Eckart potential $^{[17]}$ whose spectrum can exactly be
determined. For this
purpose we transform the Klein--Gordon equation in the form of the Schr\"{o}%
dinger-like equation, because there are many papers to tackle the
problem in the framework of Schr\"{o}dinger equations. The
eigenvalues and eigenfunctions of the Eckart potential are
obtained in terms of the SUSYQM.

\section{SUSYQM approach to bound state solution}

Generally, the s-wave Klein--Gordon equation with scalar potential
$S(r)$
and vector potential $V(r)$ can be written $^{[12,17]}$ $(\hbar =1,c=1)$%
\begin{equation}
\left\{ \frac{d^{2}}{dr^{2}}+\left[ E-V(r)\right]
^{2}-[M+S(r)]^{2}\right\} f(r)=0,  \label{e1}
\end{equation}%
where $E$ is the energy, and $M$ is the mass of the particle.
Indeed, the original wavefunction can be expressed as
$R(r)=f(r)/r$. We consider the
standard Eckart potential in the form%
\begin{equation}
V(r)=V_{1}\text{sech}^{2}(\alpha r)-V_{2}\tanh (\alpha r).
\label{e2}
\end{equation}%
When we consider the case that the vector potential and the scalar
potential
are equal, i.e. $V(r)=S(r)$, Eq.\thinspace (1) becomes a well-known Schr\"{o}%
dinger equation
\begin{equation}
\left\{ \frac{d^{2}}{dr^{2}}+(E^{2}-M^{2})-2(E+M)[V_{1}\text{sech}%
^{2}(\alpha r)-V_{2}\tanh (\alpha r)]\right\} f(r)=0,  \label{e3}
\end{equation}%
with the effective potential
\begin{equation}
V_{\text{eff}}(r)=2(E+M)[V_{1}\text{sech}^{2}(\alpha r)-V_{2}\tanh
(\alpha r)].  \label{e4b}
\end{equation}%
Then Eq. \thinspace (3) takes the form
\begin{equation}
\left\{ -\frac{d^{2}}{dr^{2}}+V_{\text{eff}}(r)\right\}
f(r)=\lambda f(r), \label{e5}
\end{equation}%
where $\lambda =E^{2}-M^{2}$ is the redefined energy parameter. In
order to solve Eq. \thinspace (5), in the framework of the SUSYQM,
we introduce the following ground-state wave function
\begin{equation}
f_{0}(r)=N\exp \left[ \int W(r)dr\right] ,  \label{e6}
\end{equation}%
where $N$ is a normalization constant, and $W(r)$ refers to a
super-potential. Substituting Eq. \thinspace (6) into Eq.
\thinspace (5), we obtain
\begin{equation}
W^{2}(r)-W^{\prime }(r)=V_{\text{eff}}(r)-\lambda _{0}, \label{e7}
\end{equation}%
where $\lambda _{0}$ is the ground-state energy, and Eq.
\thinspace (7) is a nonlinear Riccati equation which gives the
wavefunction of the system.

Our task is now, to obtain the super potential $W(r),$ which is
helpful to express the super partner potentials $V_{+}(r)$ and
$V_{-}(r).$ After some straightforward calculation, we obtain the
super potential $W(r)$, which can
be written as%
\begin{equation}
W(r)=A-B\tanh (\alpha r),  \label{e8}
\end{equation}%
where $A$ and $B$ are the constant coefficients. Notice that the
result Eq. \,(8) shows that the problem can be treated in the
framework of the SUSYQM. The

super-symmetric partner potentials are given by
\begin{equation}
V_{\pm }(r)=W^{2}(r)\pm W^{\prime }(r).  \label{e9b}
\end{equation}%
Substituting Eq. \,(8) into Eq. \,(9), we obtain the following
partner potentials
\begin{subequations}
\begin{eqnarray}
V_{+}(r) &=&A^{2}+B^{2}-B(B+\alpha )\text{sech}^{2}(\alpha
r)-2AB\tanh
(\alpha r),  \label{e10a} \\
V_{-}(r) &=&A^{2}+B^{2}-B(B-\alpha )\text{sech}^{2}(\alpha
r)-2AB\tanh (\alpha r).  \label{e10b}
\end{eqnarray}%
In order to obtain $\lambda _{0}$ and the relations of $A$ and $B$ with $%
V_{1}$ and $V_{2},$ we compare Eqs.\,(3), (7), (10a),and (10b). As
a consequence, one can easily obtain the following relations
\end{subequations}
\begin{subequations}
\begin{eqnarray}
\lambda _{0} &=&A^{2}+B^{2},  \label{e11a} \\
B &=&\frac{1}{2}\left[ \alpha \pm \sqrt{8(E+M)V_{1}+\alpha
^{2}}\right],
\label{e11b} \\
A &=&\frac{2(E+M)V_{2}}{\alpha \pm \sqrt{8(E+M)V_{1}+\alpha
^{2}}}. \label{e11c}
\end{eqnarray}%
It is well known that the potentials are shape invariant., that
is, $V_{+}(r) $ has the same functional form as $V_{-}(r)$ but
different parameters except for an additive constant. shape
invariant condition can be expressed as
\end{subequations}
\begin{equation}
V_{+}(r,a_{0})=V_{-}(r,a_{1})+R(a_{1)}),  \label{e16}
\end{equation}%
where $a_{0}$ and $a_{1}$ represents the potential parameters in
the supersymmetric partner potentials, and $R(a_{1})$ is a
constant. This property permits an immediate analytical
determination of eigenvalues and eigenfunctions. It is obvious
that the potentials are invariant when the following conditions
hold
\begin{eqnarray*}
a_{0} &=&-\frac{AB}{\alpha -B}, \\
a_{1} &=&-\alpha +B, \\
R(a_{1}) &=&a_{0}^{2}+a_{1}^{2}-(A^{2}+B^{2}).
\end{eqnarray*}%
Thus, the energy eigenvalues of Hamiltonian which includes
$V_{-}(r)$
partner potential $-\frac{d^{2}}{dr^{2}}+V_{-}(r)$ are given by%
\begin{eqnarray}
\lambda _{0}^{(-)} &=&0,  \label{e17} \\
\lambda _{n}^{(-)} &=&\sum R(a_{k})=\left( -\frac{AB}{\alpha
n-B}\right) ^{2}+(-\alpha n+B)^{2}-(A^{2}+B^{2}).  \label{e18}
\end{eqnarray}%
Therefore, the complete energy spectrum are obtained by
\begin{eqnarray}
\lambda _{n} &=&\lambda _{n}^{(-)}+\lambda _{0}=\left( -\frac{AB}{\alpha n-B}%
\right) ^{2}+(-\alpha n+B)^{2},  \notag \\
n &=&1,2,3,\cdots .  \label{e19}
\end{eqnarray}%
Substituting the values of coefficients $A,$ $B,$ and $\lambda
_{0}$ into
Eq. \,(15), we obtain the required relativistic bound-state energy spectrum%
\begin{equation}
M^{2}-E_{n}^{2}=\frac{(E_{n}+M)^{2}V_{2}^{2}}{\alpha
^{2}}\frac{1}{(n+\delta )^{2}}+\alpha ^{2}(n+\delta )^{2},
\label{e20}
\end{equation}%
where the parameters $\delta $ is defined by $\delta =-\frac{1}{2}+\frac{1}{2%
}\sqrt{1+\frac{8(E_{n}+M)V_{1}}{\alpha ^{2}}}$.

The corresponding unnormalized ground-state wavefunction is
determined by Eq. \,(6),
\begin{equation}
f_{0}(r)=N\exp (-Ar)(\cosh (\alpha r))^{B/\alpha }.  \label{ee21}
\end{equation}%
By using the parameters of $A$, $B$, and $f_{0}(r)$, we obtain the
wavefunction in the form
\begin{equation}
R_{0}(r)=\frac{1}{r}[\cosh (\alpha r)]^{(p+w)}\exp (\alpha \lbrack
w-p]r)), \label{e21}
\end{equation}%
where%
\begin{eqnarray*}
p &=&\frac{1}{2}\left[ n+\delta +\frac{2(E+M)V_{2}}{\alpha ^{2}}\frac{1}{%
n+\delta }\right], \\
w &=&\frac{1}{2}\left[ n+\delta -\frac{2(E+M)V_{2}}{\alpha ^{2}}\frac{1}{%
n+\delta }\right].
\end{eqnarray*}%
The unnormalized wavefunction of the related Hamiltonian can be
obtained by a similar mathematical procedure presented in
Ref.\,[11]. By using this method, the wavefunction can be
expressed in terms of the Jacobi polynomial as
\begin{equation}
R(r)=\frac{1}{r}[\cosh (\alpha r)]^{(p+w)}\exp [\alpha
(w-p)r]\times P_{n}^{-2p,-2w}[-\coth (\alpha r)].  \label{e22}
\end{equation}%
This equation gives the required wavefunction of the standard
Eckart potential with the Klein--Gordon equation.

\section{Conclusions}

In summary, we have discussed the exact solution of the
Klein--Gordon equation including the equal scalar and vector
Eckart potentials by using the SUSYQM. We have shown that both the
eigenvalues and eigenfunctions of the Klein--Gordon equation can
be obtained in the closed form for the Eckart potential. Finally,
we emphasize that the method discussed here can be generated for
other potentials.


\begin{thebibliography}{99}
\bibitem{a} Dirac P A M 1927 \textit{Proc. Roy. Soc. London} A \textbf{\ 114}
243

\bibitem{b} Ko\c{c} R and Koca M 2005 \textit{Mod. Phys. Lett. } A \textbf{20%
} 911

\bibitem{1} Hou C F, Sun X D, Zhou Z X and Y Li 1999 \textit{Acta Phys. Sin.}
\textbf{48} 385 (in Chinese)

\bibitem{2} Hou C F and Zhou Z X 1999 \textit{Acta Phys. Sin.} (Overseas
Edition) \textbf{8} 561

\bibitem{3} Chen C 1999\textit{\ Acta Phys. Sin. }\textbf{48} 385 (in
Chinese)

\bibitem{4} Chen C et al 2003 \textit{Acta Phys. Sin.} \textbf{52} 1579 (in
Chinese)

\bibitem{5} Qiang W C 2002\textit{\ Chin. Phys.} \textbf{11} 757

Qiang W C 2003 \textit{Chin. Phys. }\textbf{12} 1054

\bibitem{6} Hu S Z and Su R K 1991 \textit{Acta Phys. Sin. }\textbf{40} 1201
(in Chinese)

\bibitem{7} E\u{g}rifes H, Demirhan D and B\"{u}y\"{u}kk\i l\i \c{c} F 1999
\textit{Phys. Ser.} \textbf{60} 195

\bibitem{8} Jia C S et al 2003 \textit{Phys. Lett.} A \textbf{311} 115

\bibitem{9} Yi L Z, Diao Y F, Liu J Y and Jia C S 2004 \textit{Phys. Lett. }
A\textbf{333} 212

\bibitem{10} Dominguez-Adame F 1989 \textit{Phys. Lett.} A \textbf{136} 175

\bibitem{11} Chen G 2004 \textit{Phys. Scripta} \textbf{69} 257

Chen G, Chen Z D and Lou Z M 2004 \textit{Phys. Lett.} A
\textbf{331} 374

\bibitem{cao} Cao Z Q et al 2001 \textit{Phys. Rev. } A \textbf{63 } 0544103

\bibitem{he} He Y, Cao Z Q and Shen Q S 2004 \textit{Phys. Lett. } A\textbf{%
\ 326} 315

\bibitem{liang} Liang Z et al 2005 \textit{Chin. Phys. Lett. }\textbf{22}
2465

\bibitem{12} Greiner W 2000 \textit{Relativistic Quantum Mechanics} 3rd edn
(Berlin: Springer)
\end{thebibliography}
\end{document}